\documentclass[manuscript]{aastex}

\usepackage{longtable}
\usepackage{graphicx}

\shorttitle{The Absolute Magnitude Distribution of the Kuiper Belt's Scattering Objects}
\shortauthors{Shankman et al.}

\begin{document}


\title{OSSOS II: A Sharp Transition in the Absolute Magnitude Distribution of the Kuiper Belt's Scattering Population}


\author{C. Shankman,}
\affil{Department of Physics and Astronomy, University of Victoria, Elliott Building, 3800 Finnerty Rd, Victoria, British Columbia V8P 5C2, Canada}
\email{cshankm@uvic.ca}

\author{JJ. Kavelaars}
\affil{National Research Council of Canada, Victoria, BC, Canada}

\author{B. J. Gladman}
\affil{Department of Physics and Astronomy, The University of British Columbia, Vancouver, BC, Canada}

\author{M. Alexandersen}
\affil{Department of Physics and Astronomy, The University of British Columbia, Vancouver, BC, Canada}
\affil{Institute for Astronomy and Astrophysics, Academia Sinica, Taiwan}

\author{N. Kaib}
\affil{Department of Terrestrial Magnetism, Carnegie Institution for Science, Washington, DC, USA}
\affil{HL Dodge Department of Physics \& Astronomy, University of Oklahoma, Norman, OK, USA}

\author{J.-M. Petit}
\affil{Institut UTINAM, CNRS-Universit\'{e} de Franche-Comt\'{e}, Besan\c{c}on, France}

\author{M. T. Bannister}
\affil{Department of Physics and Astronomy, University of Victoria, Victoria, BC, Canada}
\affil{National Research Council of Canada, Victoria, BC, Canada}

\author{Y.-T. Chen}
\affil{Institute for Astronomy and Astrophysics, Academia Sinica, Taiwan}

\author{S. Gwyn}
\affil{National Research Council of Canada, Victoria, BC, Canada}

\author{M. Jakubik}
\affil{Astronomical Institute of the Slovak Academy of Sciences, Tatransk\'{a} Lomnica, The Slovak Republic}

\and

\author{K. Volk}
\affil{University of Arizona, Tucson, Arizona}




\begin{abstract}

We measure the absolute magnitude, $H$, distribution, $dN(H) \propto 10^{\alpha H}$ of the scattering Trans-Neptunian Objects (TNOs) as a proxy for their size-frequency distribution. 
We show that the $H$-distribution of the scattering TNOs is not consistent with a single-slope distribution, but must transition around $H_g \sim 9$ to either a knee with a shallow slope or to a divot, which is a differential drop followed by second exponential distribution.
Our analysis is based on a sample of 22 scattering TNOs drawn from three different TNO surveys --- the Canada-France Ecliptic Plane Survey \citep[CFEPS,][]{petit11}, \citet{alexandersen15}, and the Outer Solar System Origins Survey \citep[OSSOS,][]{bannister15}, all of which provide well characterized detection thresholds --- combined with a cosmogonic model for the formation of the scattering TNO population.  
Our measured absolute magnitude distribution result is independent of the choice of cosmogonic model. 
Based on our analysis, we estimate that number of scattering TNOs is (2.4-8.3)$\times 10^5$ for $H_r < 12$.
A divot $H$-distribution is seen in a variety of formation scenarios and may explain several puzzles in Kuiper Belt science.
We find that a divot $H$-distribution simultaneously explains the observed scattering TNO, Neptune Trojan, Plutino, and Centaur $H$-distributions while simultaneously predicting a large enough scattering TNO population to act as the sole supply of the Jupiter-Family Comets.

\end{abstract}





\section{Introduction}

There are many important unanswered questions about the formation of the Solar System. For example: 
What were the conditions of the initial accretion disk and how long did the planetessimals grind collisionally? 
These questions cannot be answered by direct observation of the phenomena. 
Instead, using signatures implanted in its small-body populations, we infer the formation history from the present state of the Solar System. 
The size-frequency distribution of small-body populations is shaped by the formation physics (large-sized objects) and the collisional history (small-size objects) of the population, and is thus a key signature of the history of the population (for a review see \citealt{Leinhardt2007}).
Here we study the absolute magnitude distribution of the scattering Trans-Neptunian Objects (TNOs), examining an important piece of the complex TNO puzzle.

TNOs are too small (unresolved), cold, and distant \citep{stansberry08} to allow for direct measurement of their sizes; instead, measurements of the luminosity function have been used to probe the size-frequency distribution of TNOs (see \citealt{petit08} for a review). 
Steep slopes of 0.8-1.2 have been measured for the luminosity functions of bright objects over spans of a few magnitudes (\citealt{gladmankavelaars97}; \citealt{gladman01}; \citealt{bernstein04}; \citealt{fraserkavelaars08}; \citealt{fuentesholman08}) and these steep slopes have been shown to break in the dynamically ``hot" populations at smaller size (\citealt{bernstein04}; \citealt{fraserkavelaars08}; \citealt{fuentesholman08}). 
While measuring the luminosity function has found strong features, this approach introduces uncertainties for assumptions of the size, albedo and radial distributions, which are often not well constrained in sky surveys for TNO discovery that are sensitive to faint sources.

 A more direct approach to probe the underlying size-frequency distribution is to measure the absolute magnitude distribution, or $H$-distribution. 
Measuring the $H$-distribution removes observation distance dependencies and only requires an albedo measurement to be converted into size. 
The $H$-distribution approach has been used to probe the size-frequency distribution of the TNO populations (e.g. see \citealt{gladman12}; \citealt{shankman13}; \citealt{adams14}; \citealt{fraser14}; \citealt{alexandersen15}) and here we use the $H$-distribution to probe the size-frequency distribution of the scattering TNOs.

The scattering TNOs are the best Trans-Neptunian population to use to measure the small-size $H$ or size distribution.
Because of their interactions with the giant planets, scattering TNOs come in the closest to the Sun of any TNO sub-population. 
As TNOs are discovered in reflected light and in flux-limited surveys, the best way to detect smaller objects is to have them be closer-in. 
Because of their close distances ($d$ down to 20-30 AU), a 4m class telescope can detect scattering TNOs down to $H_r~\sim$~12, which is past the observed break in the TNO hot populations seen at $H_r \sim 8$ (\citealt{shankman13}; \citealt{fraser14}). 
The scattering TNOs have smaller pericenters allowing flux-limited surveys to probe to small sizes in this population, providing an accessible sample of TNOs that cross the size range where the size-distribution appears to transition from steep (large objects) to shallow (small objects) slopes.
Here we present a measure of the $H$-distribution, a proxy for the size-frequency distribution, of the largest bias-understood sample of scattering TNOs crossing the transition (see Section~\ref{observations}).

The measured $H$-distribution of the scattering TNOs acts as a useful proxy for many other dynamically hot populations in the outer Solar System, and even beyond the Kuiper Belt, as these populations may share a common evolution. 
Several dynamical models posit that many or all TNOs formed closer-in to the Sun and were later scattered out to their current orbits \citep{gomes03,levison08,batyginbrown10,nesvorny15} during an instability phase with the giant planets. 
This would have depleted all Trojan populations of the giant planets and thus all current Neptune Trojans must have been captured from the scattering TNOs post-instability.
The Neptune Trojan population is of particular interest for its observed lack of small-sized objects \citep{sheppardtrujillo10b}, which is incompatible with a steep single-slope size-distribution.
These hot populations would share a common ``frozen-in" size-frequency distribution, formed pre-instability, as the number densities in the outer Solar System make collisions improbable.
The scattering TNOs present an opportunity to measure the small-size end of the frozen-in TNO size-frequency distribution which may be shared by the Neptune Trojans and other TNO hot populations.

To date there have been three measurements of the scattering TNO luminosity function or $H$-distribution. 
(1) With a scattering TNO sample of 4, \citet{trujillo00} do not measure the slope of the size-distribution directly, but find that slopes of $\alpha = 0.4 $ and 0.6 are not rejectable for their sample. 
(2) \citet{shankman13}, with a sample of 11 scattering TNOs, reject a single slope $H$-distribution and require a break in the $H$-distribution around $H_g$ = 9 (diameter, $D \sim 100$ km for 5\% albedo), confirming the need for a transition in the TNO hot populations. \citet{shankman13} argue in favor of a divot $H$-distribution, finding that the population of scattering TNOs with a divot $H$-distribution is numerous enough to be the source for the Jupiter Family Comets (JFCs). 
(3) \citet{adams14}, using a sample of 23 objects that includes scattering TNOs and the so-called Hot Classicals, measure the pre-break (large size) slope, finding a steep slope of $\alpha = 1.05$.
The \citet{adams14} sample includes multiple dynamical TNO populations and is thus not directly comparable to this analysis which only measures the scattering TNO population.
\citet{adams14} compare the scattering TNO $H$-distribution to their measured Centaur slope of $\alpha =$ 0.42, as the Centaurs could be supplied by the scattering TNOs. 
The \citet{adams14} Centaur sample of 7 objects only contains one object that is brighter than the break magnitude in the hot populations, and thus they measure the faint slope with no lever arm on the form of the transition at the break.
In this work we measure the $H$-distribution of the scattering TNOs, extending the sample and analysis used in \citet{shankman13}; this work provides stronger and more robust constraints on the form of the scattering TNO $H$-distribution.

In Section~\ref{observations} we discuss our observations. In Section~\ref{methods} we discuss the dynamical model used, our survey simulator approach, and our statistical tests. In Section~\ref{results} we present our results. 
In Section~\ref{discussion} we consider the implications of our results for other small-body populations in the outer Solar System and finally in Section~\ref{conclusion} we provide our concluding remarks.


\section{Observations}
\label{observations}
Observing and characterizing TNOs is difficult. TNOs are distant, faint, and move relative to the background stellar field.  
Their sky density is not uniform and they are detected in reflected light over a small range of phase angles, often exhibiting a surge in brightness near opposition \citep{benecchisheppard13}. 
The choice of pointing direction, the efficiency of tracking objects (necessary for determining orbits), and survey magnitude limits add complexity to the already difficult problem of interpreting the observed samples. 
To be properly identified, a TNO must be bright enough to be detected and then must be linked in follow-up observations to establish an orbit so that the object can be classified into a TNO sub-category. 
To take a sample of observed TNOs and determine the intrinsic population requires detailed documentation of the biases inherent in the observation process (e.g. see \citealt{kavelaars08}).  
With detailed documentation of the biases, the observations can then be ``debiased'' to infer the model from the sample or, as we do here, models of the intrinsic population can be forward biased and judged in comparison to the detected sample. 
To properly combine different surveys, the biases must be well measured for all surveys. 
We emphasize that there are a variety of factors that result in the biased sample, and each must be carefully measured, or characterized. 

Here we present our analysis on a sample of 22 scattering TNOs resulting from combining the observed samples of the Canada-France Ecliptic Plane Survey (CFEPS) \citep{petit11, kavelaarsepsc11}, \citet{alexandersen15}, and the first quarter results of the Outer Solar System Origins Survey (OSSOS) \citep{bannister15}.  
These three surveys were performed and characterized with similar approaches, allowing the samples to be combined in a straightforward manner. 
Details on the observing approach and orbital classification are given in the individual survey description papers referenced above.  
From each survey we select the scattering TNOs, as classified by the classification scheme in \citet{gladman08}: a non-resonant TNO whose orbital parameters vary by $\Delta a$ of at least 1.5 AU in a 10 Myr integration is considered to be a scattering TNO. 
As the objects in our surveys are reported in two different bands, $g$ and $r$, we adopt a  $g - r$ color for the analysis. 
As we show in Section~\ref{modelChoice}, the value of $g-r$ does not cause a material change in our results. 
The observed and derived properties of the 22 scattering TNOs used in this analysis are reported in Table 1. 

CFEPS obtained characterized observations between 2003 and 2007, covering 321 deg$^{2}$ of sky around the ecliptic to $g$-band limits of ~23.5  \citep{petit11}. 
They provided a catalog of 169 dynamically classified TNOs, 9 of which are scattering, and a set of tables that provide detailed characterization of those detections. 
The initial CFEPS catalog was supplemented by a high ecliptic latitude survey carried out in 2007 and 2008 that covered 470 deg$^{2}$, extended up to 65$^{\circ}$ ecliptic latitude and found 4 scattering TNOs \citep{kavelaarsepsc11}. 
The extended survey's detection and tracking characterization is provided in \citet{petit15}.
This combined data set of 13 scattering TNOs is referred to as the CFEPS sample. 

\citet{alexandersen15} performed a 32 deg$^{2}$ survey to a limiting $r$-band magnitude of 24.6, finding 77 TNOs. 
They found two temporary Trojans, one for Uranus and one for Neptune. 
Using the SWIFT package \citep{levisonduncan94} for orbital integrations, they found that both objects ceased to be co-orbitals within $\sim1 \, \mathrm{Myr}$, after which they both rejoin the scattering population (\citealt{alexandersen13}, \citealt{alexandersen15}).
Both objects satisfy the scattering classification criterion as above.  
The survey analysis from \citet{alexandersen15} followed the same careful characterization process as used in CFEPS.  
The characterization information for this survey can be found in \citet{alexandersen15}

In the northern hemisphere fall of 2013, OSSOS searched 42 square degrees of sky, detecting 86 TNOs brighter than the survey's limiting $r'$ magnitudes of  24.04 and 24.40 (for OSSOS's E and O blocks respectively). 
Of those, 7 were found to be on scattering orbits and are included in the analysis presented here.    
For the OSSOS and \citet{alexandersen15} surveys (as distinct from the CFEPS sample) the orbital tracking observations were more frequent during the discovery year, enabling orbital classification after only two years of observing as compared to the four to five years needed for the more sparsely observed CFEPS targets.
The complete details of the OSSOS characterization can be found in \cite{bannister15}. 

Our observed sample of 22 scattering TNOs down to $H_r$ of 12 is the largest sample of scattering TNOs from characterized surveys, and the only sample that extends beyond the confirmed break in the $H$-distribution. 
This is the best sample available to probe the form of the $H$-distribution at and beyond the transition in the size-frequency distribution.

\begin{table}
\begin{center}
\begin{tabular}{  c * {9}{c}  }
\hline
\hline
\bf{Designation} &  \bf{Designation} &  \bf{a} &  \bf{q} &  \bf{i} &  \bf{d} &  \bf{m} &  \bf{H} &  \bf{Filter} &  \bf{Survey} \\ 
\bf{Internal} &  \bf{MPC} &  \bf{(AU)} &  \bf{(AU)} &  \bf{(deg)} &  \bf{(AU)} &  \bf{ } &  \bf{ } &  \bf{ } &  \bf{ } \\ 
\tableline

L4k09 &  2004~KV18 &  30.19 &  24.6 &  13.59 &  26.63 &  23.64 &  9.33 &  g &  CFEPS \\ 
L4m01 &  2004~MW8 &  33.47 &  22.33 &  8.21 &  31.36 &  23.75 &  8.75 &  g &  CFEPS \\ 
L4p07 &  2004~PY117 &  39.95 &  28.73 &  23.55 &  29.59 &  22.41 &  7.67 &  g &  CFEPS \\ 
L3q01 &  2003~QW113 &  51.05 &  26.31 &  6.92 &  38.17 &  24.0 &  8.16 &  g &  CFEPS \\ 
L7a03 &  2006~BS284 &  59.61 &  33.41 &  4.58 &  46.99 &  23.84 &  7.12 &  g &  CFEPS \\ 
L4v04 &  2004~VG131 &  64.1 &  31.64 &  13.64 &  31.85 &  24.14 &  9.09 &  g &  CFEPS \\ 
L4v11 &  2004~VH131 &  60.04 &  22.26 &  11.97 &  26.76 &  24.19 &  9.94 &  g &  CFEPS \\ 
L4v15 &  2004~VM131 &  68.68 &  20.61 &  14.03 &  22.97 &  22.0 &  8.96 &  g &  CFEPS \\ 
L3h08 &  2003~HB57 &  159.68 &  38.1 &  15.5 &  38.45 &  24.29 &  8.36 &  g &  CFEPS \\ 
HL8a1 &  2008~AU138 &  32.39 &  20.26 &  42.83 &  44.52 &  22.93 &  6.29 &  r &  CFEPS \\ 
HL8n1/Drac &  2008~KV42 &  41.53 &  21.12 &  103.45 &  31.85 &  23.73 &  8.52 &  r &  CFEPS \\ 
HL7j2 &  2007~LH38 &  133.9 &  36.8 &  34.2 &  37.38 &  23.37 &  7.5 &  r &  CFEPS \\ 
ms9 &  2009~MS9 &  348.81 &  11.0 &  68.02 &  12.87 &  21.13 &  9.57 &  r &  CFEPS \\ 
mal01 &  2011~QF99 &  19.09 &  15.72 &  10.81 &  20.3 &  22.57 &  9.57 &  r &  A14 \\ 
mah01 &  2012~UW177 &  30.06 &  22.29 &  53.89 &  22.43 &  24.2 &  10.65 &  r &  A14 \\ 
o3o01 &  2013~JC64 &  22.14 &  13.76 &  32.02 &  13.77 &  23.39 &  11.95 &  r &  OSSOS \\ 
o3e01 &  2002~GG166  &  34.42 &  14.12 &  7.71 &  23.29 &  21.5 &  7.73 &  r &  OSSOS \\ 
o3o36 &  2013~JQ64  &  48.79 &  22.38 &  34.88 &  57.34 &  23.73 &  6.09 &  r &  OSSOS \\ 
o3o16 &  2013~JP64 &  57.44 &  32.35 &  13.7 &  35.68 &  23.92 &  8.34 &  r &  OSSOS \\ 
o3o17 &  2013~JR64  &  77.56 &  35.64 &  10.46 &  35.81 &  24.31 &  8.71 &  r &  OSSOS \\ 
o3e11 &  2013~GZ136  &  86.74 &  33.89 &  18.36 &  36.85 &  23.6 &  7.86 &  r &  OSSOS \\ 
o3o14 &  2013~JO64  &  143.35 &  35.13 &  8.58 &  35.46 &  23.54 &  8.0 &  r &  OSSOS \\ 

\tableline
\end{tabular}
\end{center}
\caption[]{\label{Table:sTNO} The combined scattering TNO samples from CFEPS, OSSOS and \citet{alexandersen15}. The magnitude and $H$-magnitude given are both in the listed filter. }
\end{table}


\section{Methods}
\label{methods}

Our observationally biased sample of scattering TNOs can be used to explore the characteristics of the intrinsic scattering population via a model comparison.
Through the process of characterization,  each input survey provides a careful estimate of the detection and tracking bias that is present in the detected sample.
Rather than de-bias our observed sample into an estimate of the intrinsic population, we forward bias intrinsic orbital models of the scattering TNOs and compare them with the observed sample.  
We forward bias a model of scattering TNOs using our Survey Simulator \citep{jones06, petit11}.  
The resulting biased model of the intrinsic population is then tested  by comparison to the detected sample. 
Each model is the combination of an orbital model paired with an $H$-magnitude distribution in a specific filter, a color conversion distribution, and light curve effects. 
We test the joint model by comparing orbital parameters (semi-major axis, inclination, pericenter) and observed parameters ($H$-mag, distance at detection, magnitude at detection) of the simulated detections against our observed sample via the Anderson-Darling (AD) test (see Section~\ref{stats}). 
We show in Section~\ref{modelChoice} that the rejection of this combined model constitutes a rejection of the $H$-magnitude distribution and we thus are able to determine the $H$-magnitude distribution of the scattering TNOs.
This approach introduces orbital and color model dependencies. 
We show in Section~\ref{modelChoice} that our analysis is not sensitive to the choice of orbital model or color distribution. 

All tools required to perform this analysis are available at: http://dx.doi.org/10.5281/zenodo.31297

\subsection{Survey Simulator}
\label{SS}
The Survey Simulator determines whether a given object would have been detected and tracked by one of our surveys. 
The simulator is given a list of survey pointings and the detection efficiency for each pointing in order to perform a simulated survey. 
A randomly selected model object, with an assigned $H$-magnitude and color, is tested for detection by the survey simulator. 
The simulator first checks that the object is bright enough to have been seen in any of our surveys' fields, 
then checks that the object was in a particular field, and that it was bright enough to be detected in that field.
Based on a model object's simulator-observed magnitude and the field's detection and tracking efficiencies, model objects are assessed for ``observability".
The survey simulator reports the orbital parameters, the specified $H$-magnitude, and color conversion for all orbital model objects determined to have been ``detected" and ``tracked".
The object's ``observed" magnitude, and corresponding $H$-magnitude, which includes accounting for measurement uncertainties, are also reported.
Using the Survey Simulator, we produce a statistically large model sample that has been biased in a way that matches the biases present in our observed survey sample.
This large Survey Simulator produced model sample is then compared directly to the detected sample. 

\subsection{Models}
\label{models}
\begin{figure}[!h]
\includegraphics[width=1.0\textwidth]{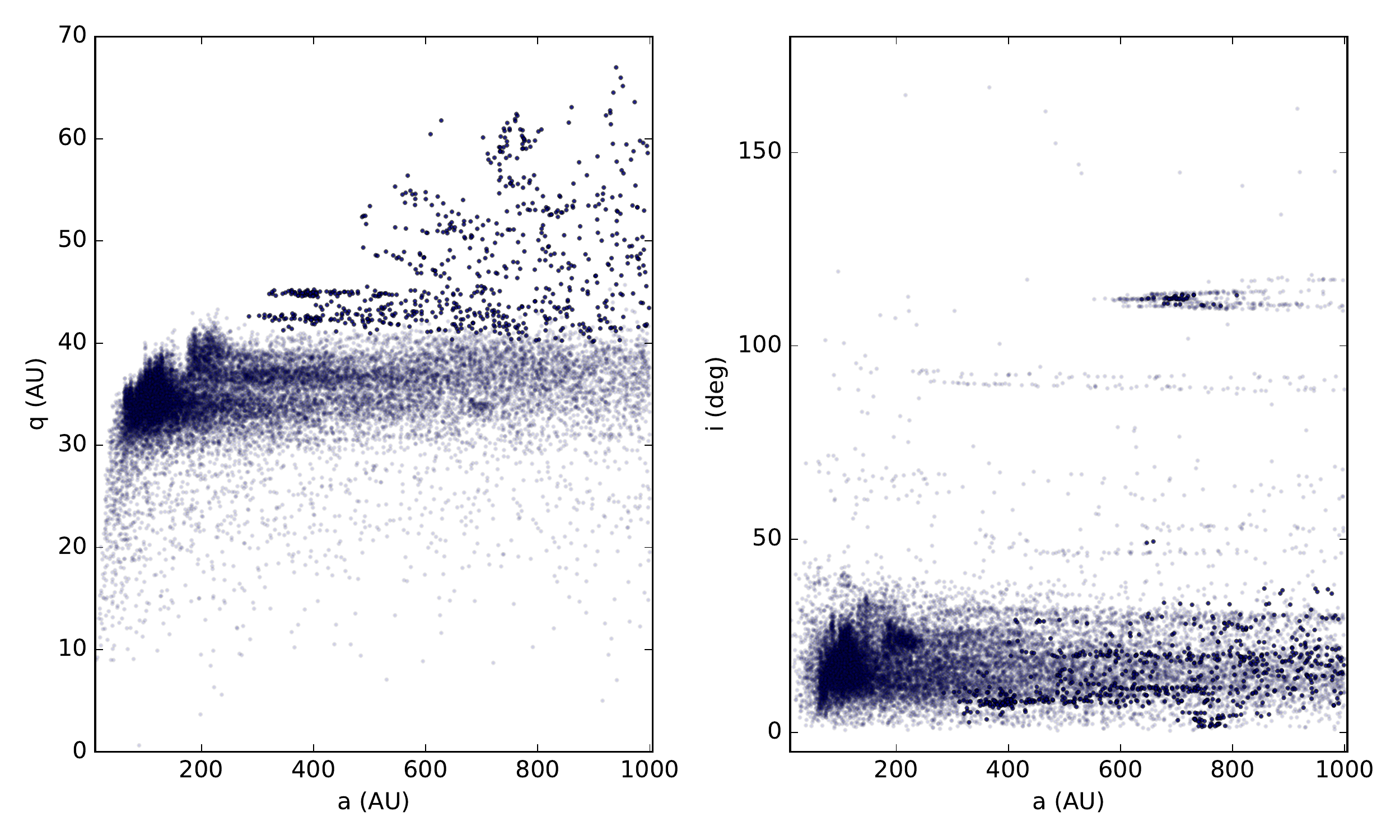}
\caption{ 
Scatter plots of pericenter $q$ vs $a$ and $i$ vs $a$ for the modified \citet{kaib11b} orbital model, sliced as $a < 1000$ AU, and $q$ $< 200$ AU. Each point represents one object in the simulation. 
Points have transparency of 0.1 to highlight densities in the model. Some points appear solidly dark as they were cloned in place at the end-state of the simulation to balance that 
they were not cloned at any other point in the simulation (for details see \citealt{kaib11b}).
}
\label{fig:KRQmodel}
\end{figure}

In order to carry out a simulated survey, one requires an orbital model for the objects being ``observed". 
For our model we select out the scattering TNOs from a modified version of the \citet[KRQ11]{kaib11b} orbital model of the TNO population.
The KRQ11 model (see Figure~\ref{fig:KRQmodel}) is the end-state of a dynamical simulation of the evolution of the Solar System that includes the gravitational effects of the giant planets, stellar passages and galactic tides.
The simulation begins with an initial disk of massless test particles between semi-major axis  $a = 4$ AU and $a = 40$ AU following a surface density proportional to $a^{-3/2}$, eccentricities, $e$ $<0.01$ and inclinations, $i$, drawn from $\mathrm{sin(}i\mathrm{)}$ times a gaussian, as introduced by \citet{brown01}. 
The giant planets are placed on their present-day orbits (see Section~\ref{modelChoice} for a discussion on the effects of the planet configuration and how it does not affect our result), the stellar neighbourhood is modeled assuming the local stellar density, and the effects of torques from galactic tides are added (for more detail, see \citealt{kaib11b}). This system is then integrated forward in time for 4.5 Gyr, resulting in a model for the current state of bodies in the outer Solar System.
The resulting orbital distribution is then joined with a candidate $H$-distribution, and a TNO color distribution derived from \citet{petit11}.   
This joint orbit, $H$-distribution and color distribution model forms the input for the Survey Simulator.


\citet{shankman13} demonstrate that the inclinations in KRQ11 are too low to match the observed scattering TNOs; the model's assumed initial inclination distribution is too ``cold". 
A model with a ``hotter" initial $i$ distribution (with gaussian width $\sigma = 12^{\circ}$) was created and dynamically evolved forward for the age of the Solar System. 
We continue to use this modified KRQ11 as our orbital model in this analysis. 
In Section~\ref{modelChoice} we discuss the effect that the choice of model has on our analysis.

We use the modified KRQ11 orbital model as a representation of the current-day population of scattering objects in the Solar System in order to perform a simulated survey of scattering TNOs. 
We select out the scattering TNOs from the orbital model (i.e. those with $\Delta a \geq 1.5$ AU in 10 Myr) in plausibly observable ranges ( $a < 1000$ AU, pericenter $q$ $< 200$ AU), which results in a relatively small number of objects.
To account for the finite size of the model, we draw objects from the model and add a small random offset to some of the orbital parameters. 
The scattering TNOs don't have specific orbital phase space constraints (unlike the resonances which are constrained in $a$, $e$, and resonant angles) and thus we can better sample the space the model occupies by slightly adding this small random offset.
This extends the model beyond the set of TNOs produced in the original run ($\sim$29k for above $a$, $q$ slices) that was necessarily limited by computation times. 
We resample a, q, and i by randomly adding up to $\pm$ 10 \%,  $\pm$ 10 \%, and $\pm$ $1^{\circ}$, respectively, to the model-drawn values. 
We also randomize the longitude of the ascending node, the argument of pericenter and the mean anomaly of each model object.
This modified KRQ11 orbital model, resampled to increase its utility, is used as the input orbital model for the scattering TNOs.

\subsection{$H$-magnitude Distribution} 
\label{Hmag}

A variety of forms have been used to try to match the observed magnitude or $H$-magnitude distributions of various TNO populations. 
Single slopes (e.g. \citealt{jewitt98}; \citealt{gladman01}; \citealt{fraserkavelaars08}; \citealt{gladman12}; \citealt{adams14}), knees (\citealt{fuentesholman08}; \citealt{fraserkavelaars09}; \citealt{fraser14}), knees with smooth rollovers (\citealt{bernstein04}), and divots (\citealt{shankman13}; \citealt{alexandersen15}) have all been proposed. 
Here we present a generalized form of the $H$-magnitude distribution for testing, which in limiting cases becomes either a knee or a single-slope.


\begin{figure}[h!]
\includegraphics[width=0.75\textwidth]{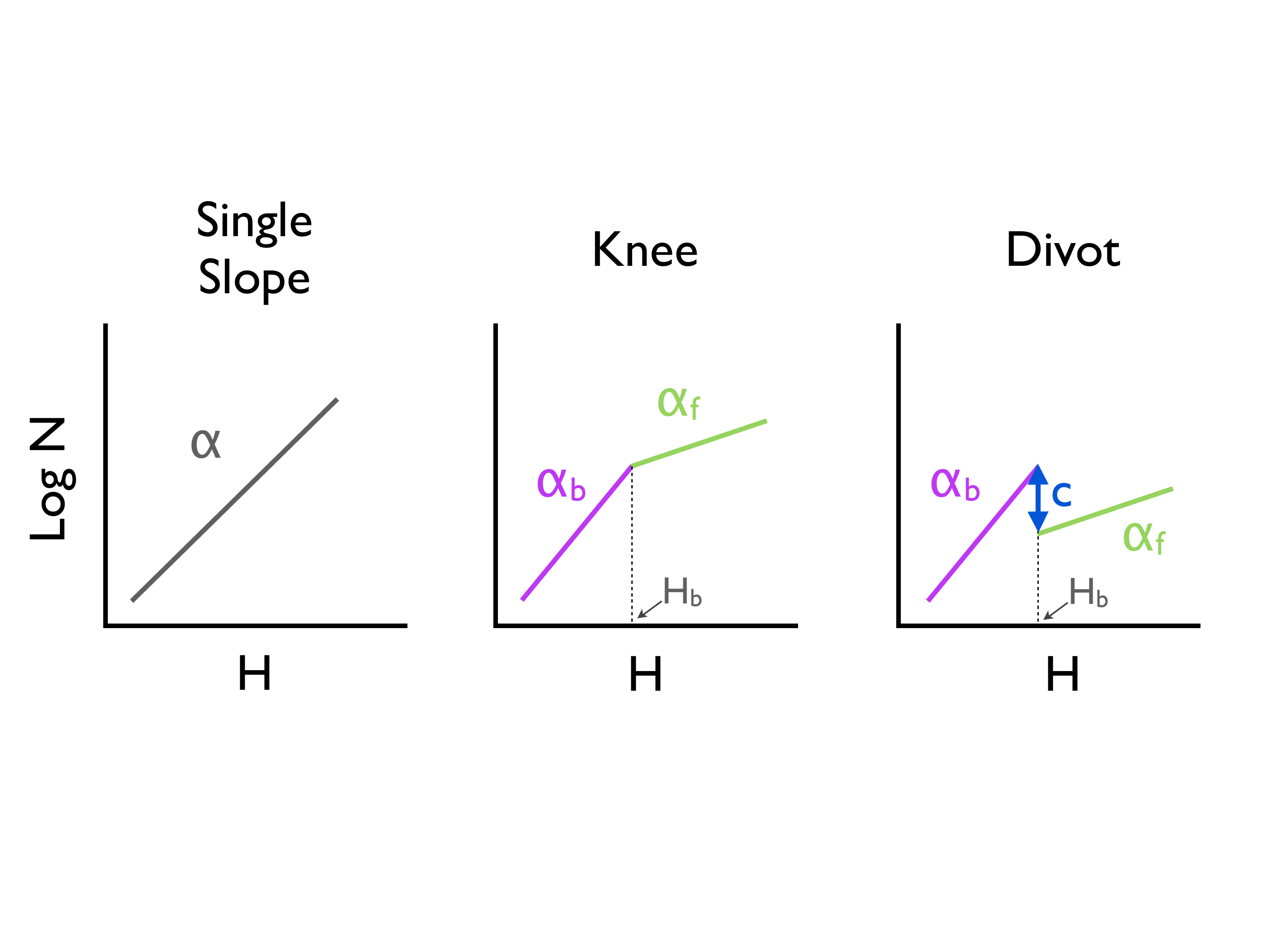}
\caption{ A schematic of the differential forms of the three $H$-distribution cases we test: single slope, knee, and divot. A single slope is parameterized by a logarithmic slope $\alpha$. A knee is parameterized by a bright slope $\alpha_b$, a faint slope $\alpha_f$, and a break location $H_{b}$. A divot is parameterized as a knee, but with a contrast $c \ ( \geq 1 )$, which is the ratio of the differential number of objects right before the divot to the number just after the divot, located at $H_{b}$. }
\label{fig:Hschem}
\end{figure}


We characterize the $H$-distribution by four parameters: a bright (large-size) slope, $\alpha_b$, a faint (small-size) slope, $\alpha_f$, a break absolute magnitude, $H_b$, and a contrast, $c$, that is the ratio of the differential frequency of objects just bright of the break to those just faint of the break.
Depending on the parameters, our $H$-distribution takes one of three forms: single-slope, knee, or divot (schematic shown in Figure~\ref{fig:Hschem}). 
We model the transition in the $H$-distribution as an instantaneous break, as the sample size of our observation does not merit constraining the form of a potential rollover.
All $H$-distributions and values given in this work are presented in $H_r$ unless otherwise specified.
Our formulation of the $H$-distribution allows for the testing of the proposed $H$-distribution from a single framework.

In the literature, the single slope distribution has been referred to as a single power-law distribution because it corresponds directly to the theorized distributions of diameters, $D$, which is a power-law to an exponent, $q$
: $\frac{\mathrm{d}N}{\mathrm{d}D} \propto D^{-\mathrm{q}}$.
This distribution is convertible to absolute magnitude and parameterized by a logarithmic ``slope" $\alpha$: $\frac{\mathrm{d}N}{\mathrm{d}H} \propto 10^{\alpha H}$, with $q$~$=5\alpha+1$.
$H$-distributions can be mapped to $D$-distributions, with an albedo, providing an observable way to probe size-distributions.

In order to create a synthetic $H$-distribution sample with a transition, one randomly samples from the two single-slope distributions (bright and faint ends), choosing which to sample from according to the fraction of the total distribution each section comprises. 
If the bright end of the distribution accounts for 60\% of the whole distribution, then when randomly drawing objects, 60\% of the time they should be drawn from a single-slope $H$-distribution corresponding to the bright distribution, and thus 40\% of the time they should be drawn from the faint end distribution. 
The ratio of the number of objects in the bright end of the distribution to the total distribution depends on the two slopes, the contrast, the break magnitude, and the faintest magnitude, $H_{faintest}$; no normalization constants or knowledge of population size are required. 
The ratio can be calculated with (see \citealt{shankman12} for a derivation):  
\begin{equation}
{\frac{N_{bright}}{N_{total}} = \left(1 + \: \frac{\alpha_b}{\alpha_f}\: \frac{1}{c}\: \left(10^{\alpha_f\left(H_{faintest} -H_b\right)} - 1\right)\right)^{-1} }
\end{equation} 

The $H$-distribution has four parameters: $\alpha_b$, $\alpha_f$, $H_{break}$, and $c$; with arguments from other TNO populations, we fix two of these parameters. 
Our sample of large-size objects is not large enough to measure the large-end slope, $\alpha_b$.
Motivated by other hot TNO populations, we fix $\alpha_b$ = 0.9, which matches our bright-end sample and is consistent with the slopes found for the hot Classical belt (\citealt{fuentesholman08}; \citealt{fraserkavelaars09}; \citealt{petit11}), the aggregated hot population (\citealt{adams14}; \citealt{fraser14}), the 3:2 resonators \citep{gladman12, alexandersen15} and the pre-transition scattering plus Hot Classical TNOs \citep{adams14}. 
As in \citet{shankman13}, we fix the break location to $H_r$ = 8.3 (corresponding to $H_g$ = 9.0 and $D \approx$ 100 km for 5\% albedo). 
A single-slope $H$-distribution of  $\alpha$ = 0.9 to the break magnitude of $H_r = 8.3$ is not rejectable at even the 1-$\sigma$ level by our sample, and thus provides good agreement to our observations.
A steep slope and break at $H_r$ = 8.3 is consistent with the \citet{adams14} scattering and Hot Classical TNO $H$-distribution with $\alpha = 1.05$ measured down to $H_r \sim 6.7$ that did not find a break, and the hot population breaks of $H_r$ = 8.4 and $H_r$ = $7.7^{+1.0}_{-0.5}$ found by \citet{fraser14}. 
Our detected sample near the transition is not large enough to constrain the exact position of the break; moving the break location by several tenths of magnitude does not affect the conclusions of this work. 
Having fixed two parameters, we are left $\alpha_f$ and $c$ as free parameters to test and constrain.

\subsection{Colors and Light Curves}
\label{colors}

Because CFEPS was primarily performed in $g$-band and \citet{alexandersen15} and OSSOS were performed in $r$-band, we must account for conversion between these two filters. 
The majority of our detections (13 vs 9) are $r$-band detections and so we choose to convert everything to $r$, requiring us to adopt a $g - r$ color conversion for both the survey simulator and our detections. 
We do not have measurements of $g$ and $r$ for all of our objects, and so we adopt the CFEPS reported color conversion of $g - r = 0.7$ \citep{petit11} which is consistent with the color measurements we do have for our scattering TNOs. This conversion allows us to combine the samples from several surveys.

We convert all of our observed scattering TNOs into $r$-band with the above conversion. 
We select colors for our modeled objects from a gaussian $g - r$ distribution centered at 0.7 with a  standard deviation of 0.2 (the range seen in the CFEPS object sample with $g-r$ available). 
We test our analysis by choosing extreme values for our color conversions (i.e. 0.5 and 0.9).
The effects of these color choices do not alter the conclusions of the analysis (see Section~\ref{modelChoice}). 

While the survey simulator allows for the modelling of light curve effects, we do not model the unknown light curves of our scattering TNO sample.
Each of our detections is measured at a random phase of its light curve and the magnitude of light curve variation is small in comparison to the uncertainty in converting between detection band passes, which we show has no meaningful impact on our analysis (see Section~\ref{modelChoice}). 
We conclude that light curves do not affect our analysis.

\subsection{Statistics}
\label{stats}

We want to test whether our model (orbital distribution, $H$-magnitude distribution, and observation biases) is a good representation for our observations. 
Having generated a set of simulated detections for a candidate $H$-distribution, we then test the hypothesis that our detected sample can be drawn from the simulated detections. 
If this hypothesis is rejected, the model used to generate our simulated detections is rejected; in particular, we conclude that this is a rejection of the candidate $H$-magnitude distribution used. 
We use a variant of the Anderson-Darling statistical test \citep{andersondarling54} to assess if our observations can be drawn from our simulated detections for a candidate $H$-magnitude distribution.

The Anderson-Darling test is a variant of the Kolmogorov-Smirnov two-sample test, weighted such that there is greater sensitivity to the tails of the distribution \citep{andersondarling54}. 
The test metric can be thought of as the distance between two cumulative distributions. 
Typically, the test is used to determine if data are consistent with being drawn from a well known distribution (e.g. normal, uniform, lognormal) for which there is a look-up table of critical values for the AD metric. 
In our case, our simulated detections take the place of the well known distribution, and we have no look-up table of critical values for the metric. 
We therefore bootstrap our sample, repeating draws of 22 objects (the same as our observed sample), calculating their AD distance from their parent distribution in order to build a distribution of critical values for the simulated sample distribution. 
If the AD metric for our observed sample is in the tails (3-$\sigma$ of the bootstrap-built distribution for our simulated detections), we reject the hypothesis that our observations could have been drawn from the candidate simulated detection distribution and we thus reject the candidate $H$-distribution used.

We apply the AD test across several variables in our data set.
We have a set of observed scattering TNOs which can be characterized by their orbital parameters and their observable characteristics. 
We test our model on a combination of three orbital parameters (semi-major axis, inclination, pericenter) and three observable parameters (magnitude at detection, distance at detection, $H$-magnitude). 
We sum the AD metrics for all of our tested distributions, and use this summed metric to test for the rejection of our model. 
While bootstrapping to build up our distribution of critical values as described above, we sample out 22 objects and simultaneously compute the AD metric for each distribution, as opposed to testing the distributions independently.
This preserves the relationships between the parameters; in essence we calculate the dimensionless AD metric for a set of 22 orbits across our six observational and dynamical characteristics to build a distribution of AD values. 
We use the sum of the AD metrics, as the sum will be small if all of the distributions are in good agreement, large if one is in poor agreement or several are in moderately poor agreement, and larger still if all are in poor agreement \citep{parker15}. 
This approach has less power than rejecting based on the worst failing tested distribution, as has been commonly done in the literature, but is a more robust and multivariate approach.


\section{Results: Absolute Magnitude Distribution}
\label{results}

We test a grid of $\alpha_f$ (0.0 - 0.9) and contrast, $c$, (1 - 100) values, and a range of single-slope $\alpha$ values (0.1 to 0.9), performing a simulated survey for each candidate $H$-distribution. 
Our scattering TNO sample rejects single slopes, requiring a transition to a shallower $\alpha_f$ slope.
We determine the plausible $c$ and $\alpha_f$ values that are consistent with our observed sample. 
The depth of the added surveys and the added observation sample provide a significant improvement in constraining the $H$-distribution, with a much higher rejection power over the analysis of the CFEPS sample alone \citep{shankman13}.

\begin{figure}[h!]
\includegraphics[width=1.2\textwidth]{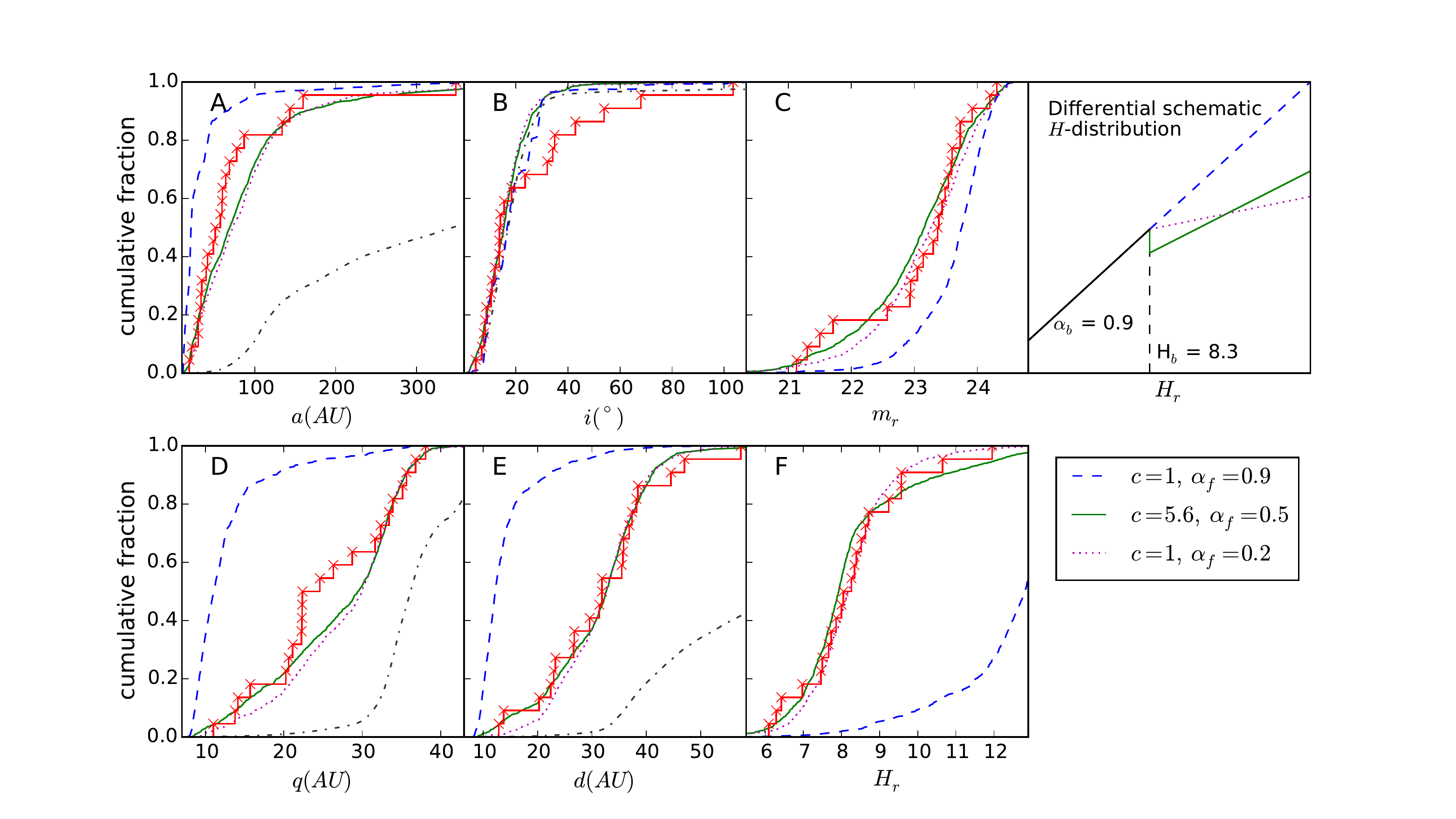}
\caption{ 
Cumulative distributions across six parameters for the intrinsic orbital model (black dash-dot), the observations (red step-function), and three candidate $H$-distributions (solid green, dotted magenta, and dashed blue). 
Panels A-F correspond to the semi-major axis, inclination, magnitude at detection in $r$-band, pericenter, distance at detection, and $H$ magnitude in $r$, respectively. 
The rightmost panel provides schematics for our three $H$-distributions: 
\textbf{(1)} our preferred ($c$, $\alpha_f$) pair (solid green; see Section~\ref{divotArguments}), 
\textbf{(2)} the $\alpha = 0.9$ single-slope distribution (dashed blue), and 
\textbf{(3)} for comparison, our knee distribution which is closest to the \citet{fraser14} hot population $H$-distribution (dotted magenta).  
}
\label{fig:cumulativeDSS}
\end{figure}

We find that there is a dearth of small objects in the detected sample compared to the number predicted by $H$-distributions with steep slopes. 
Steep slopes dramatically overpredict the number of small objects, providing strong leverage to reject such $H$-distributions. 
The blue dashed line in Figure~\ref{fig:cumulativeDSS} shows the over-prediction of a single-slope of 0.9, in particular in panels E and F. 
This is the the strongest rejection of a single-sloped $H$-distribution in a TNO sub-population.

\begin{figure}[h!]
\includegraphics[width=0.75\textwidth]{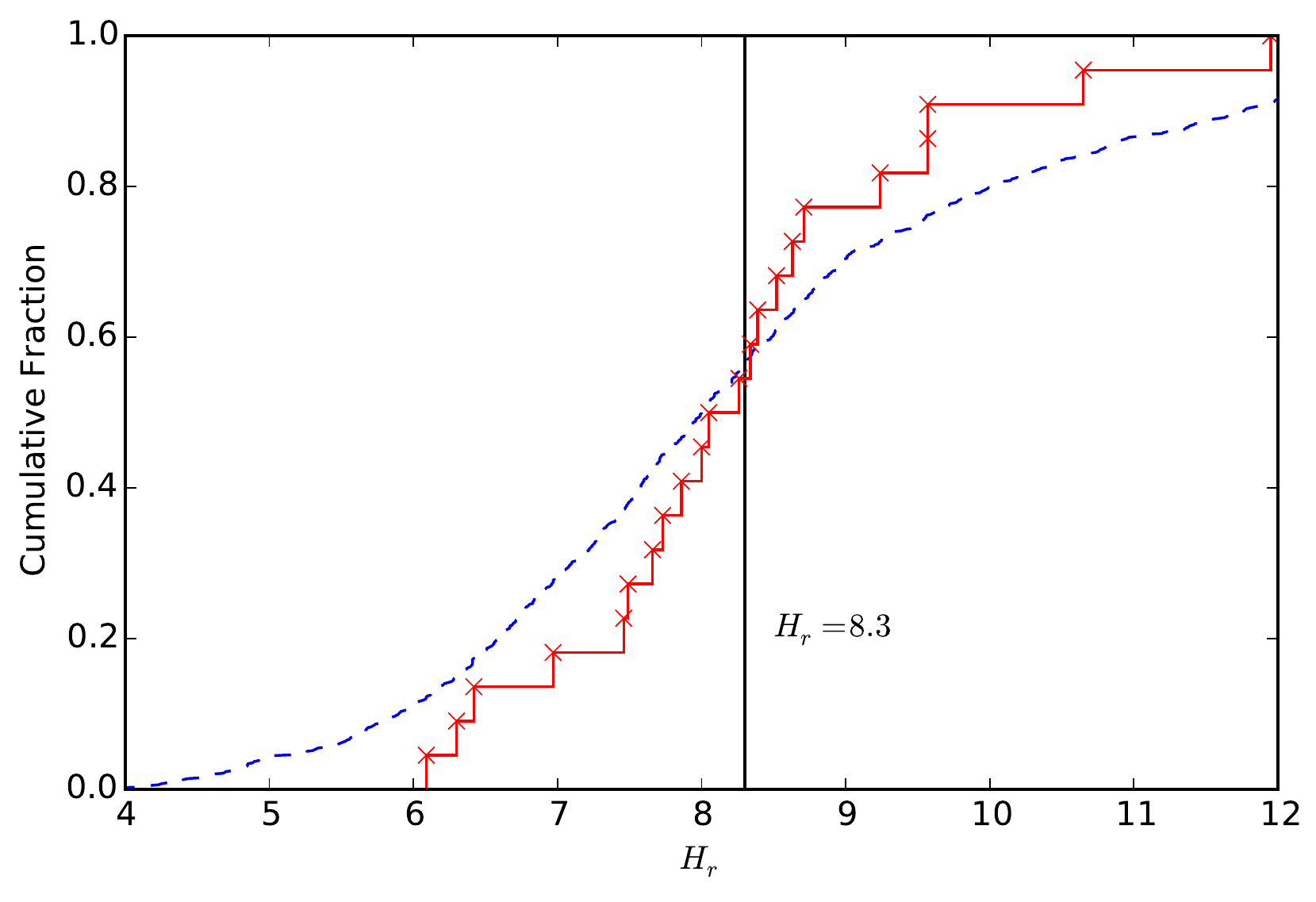}
\caption{ 
Cumulative $H$-distributions in $r$ for our observed sample (red step function) and a simulator biased single-slopes of $\alpha = 0.5$ (dashed blue).
The single-slope of 0.5 over predicts for $H < 8.3$ and under predicts for $H > 8.3$, but provides an acceptable match when measuring across a transition from steep to shallow.
}
\label{fig:cumulativeSPL05}
\end{figure}

Our sample rejects all single slopes in the range 0.1 - 0.9 at greater than 95\% significance with the exception of  $\alpha_f$ = 0.5, due to measuring a shallow slope across a transition. 
When measuring a small sample across a transition from a steep region to a shallow region, a moderate slope can average out to an acceptable match across the distribution. 
This effect can also explain the shallow slope found by \citet{fraser10}.
Testing the 0.5 $H$-distribution in pre-break and post-break slices of $H$, we find that the candidate $H$-distribution performs substantially better across the whole distribution, than in either of the slices.
Figure~\ref{fig:cumulativeSPL05} shows that the $H$-distribution for a single-slope of 0.5 does not provide a compelling match to the detected sample.
A slope of 0.5 has been ruled out by other measurements of the hot TNO populations.
A single slope of 0.5 is inconsistent with prior constraints and the non-rejection is likely an artifact arising from measuring a small sample across the transition, and therefore we do not consider a slope of 0.5 to be a plausible $H$-distribution for the scattering TNOs. 
Our analysis rejects a single slope ($c$ = 1, $\alpha_f = \alpha_b$) as the form of the scattering TNO $H$-distribution.

The KRQ11 inclination, $i$, distribution is not representative of our observed sample and is excluded from our analysis.
The KRQ11 orbital model contains too few high-$i$ objects to match our observations (see Figure~\ref{fig:cumulativeDSS}).
The intrinsic and survey simulator-observed $i$-distributions are very similar, regardless of choice of $H$-distribution (see Figure~\ref{fig:cumulativeDSS} panel B), indicating that the discrepancy comes from the intrinsic KRQ11 $i$-distribution.
The semi-major axis, $a$, distribution is strongly sensitive to the discrepant $i$-distribution and we thus exclude it from the analysis as well.
We demonstrate in Section~\ref{modelChoice} that the $m$, $q$, $d$, and $H$ distributions are not dependent on the $i$ and $a$ distributions and that our analysis is not affected by the exclusion of $a$ and $i$.

\begin{figure}[h!]
\includegraphics[width=1\textwidth]{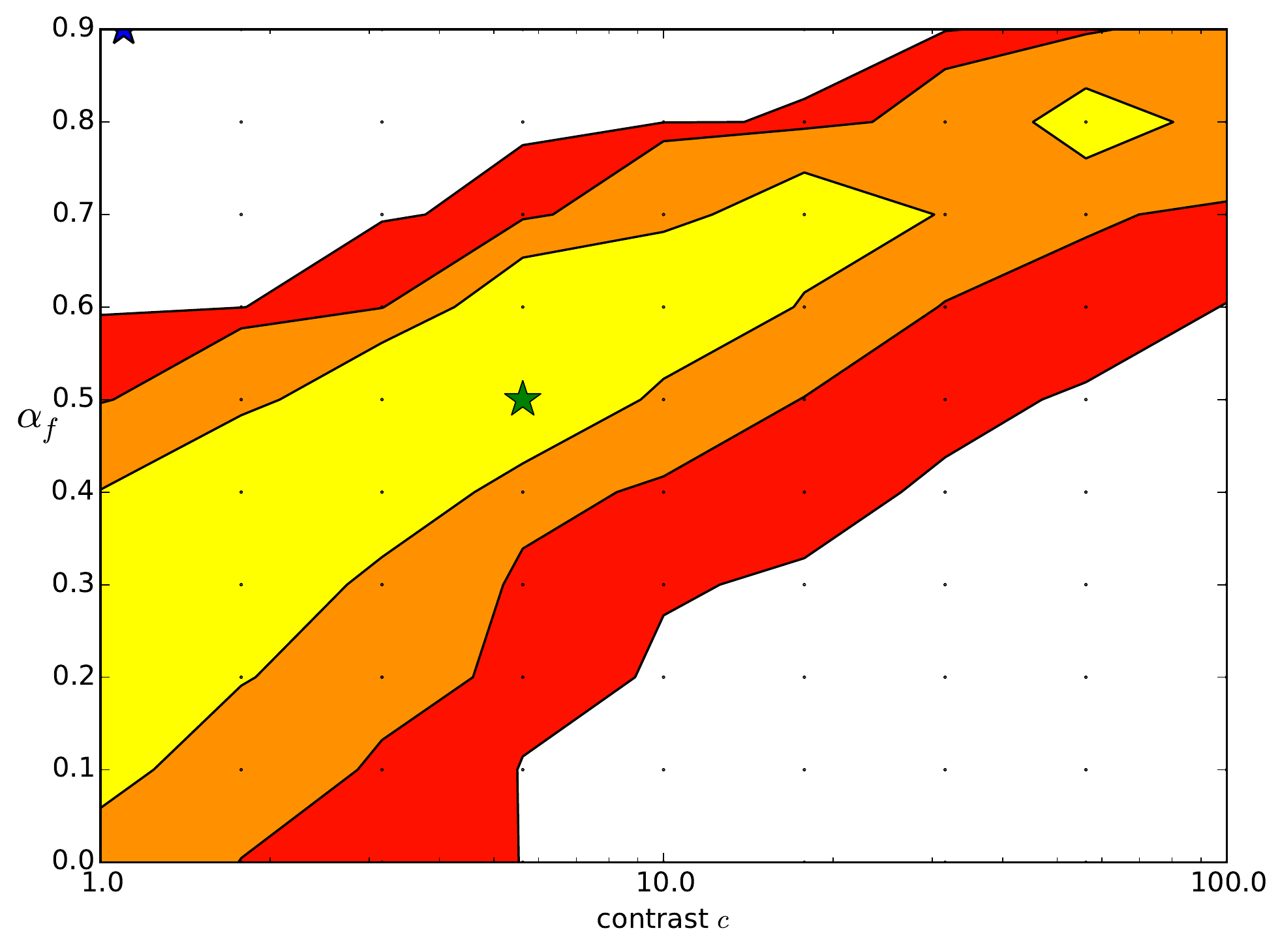}
\caption{
Contours of the rejectability for the tested $\alpha_f$ and contrast pairs. The contours represent the 1-$\sigma$, 2-$\sigma$ and 3-$\sigma$ rejectability levels with white being rejectable at $>$ 3-$\sigma$, red being rejectable at $>$ 2-$\sigma$, orange rejectable at $>$ 1-$\sigma$ and yellow rejectable at $<$ 1-$\sigma$. The green star denotes our preferred ($\alpha_f$, $c$) pair (see Section~\ref{divotArguments}), and the blue star, offset from $c =1$ for clarity, denotes the single slope of $\alpha = 0.9$
}
\label{fig:contour}
\end{figure}

We find that the scattering TNO $H$-distribution is consistent with a range of model distributions that exhibit a knee ($c = 1$) or a divot ($c > 1 $). 
In Figure~\ref{fig:contour} we present a grid of ($\alpha_f$, $c$) values that describe $H$-distributions that are consistent with our scattering TNO sample.
All models with $\alpha_f = \alpha_b$ (i.e. single-slope models) are rejected.
Figure~\ref{fig:contour} shows that models with $\alpha_f > 0.5$ and $c=1$ are rejected at or above the 3-$\sigma$ level.
Models with $c = 1$ (i.e. a knee) provide an acceptable distribution for $ 0 < \alpha_f < 0.5$.
Table~\ref{Table:collapCon} gives the non-rejectable $c$ ranges for each tested $\alpha_f$.
In Table~\ref{Table:popEst} we provide population estimates to the limit of our observed sample, $H = 12$, for divots along the 1-$\sigma$ ridge and all knee distributions not rejectable at 1-$\sigma$. 
In this $H$ range, the population estimates are relatively consistent, with the largest and smallest estimates differing by a factor of $\sim 3$.
The size of the scattering TNO population provides an observational constraint on the population's $H$-distribution; our analysis converges to $H$-distributions that imply a similar size for the intrinsic population in the size range that our survey is sensitive to.
Using only this dataset, we constrain the form of the $H$-distribution to knees and divots; knee distributions may be considered as having fewer free parameters (expressly forcing $c=1$ as special case of the model) and are thus preferred when no external factors are considered. 
Having determined the possible parameters of the $H$-distribution, in the next section we consider the implications of our result for other populations in the outer Solar System.

\begin{table}
\begin{center}
\begin{tabular}{  c  * {3}{c }  }
\hline
\hline
\bf{$\alpha_f$} &  \bf{$c$} \\ 
\tableline

0.9 &  56.2 - 100.0 \\ 
0.8 &  17.8 - 100.0 \\ 
0.7 &  15.6 - 100.0 \\ 
0.6 &   3.2 - 56.2 \\ 
0.5 &  1.0 - 31.6 \\ 
0.4 &  1.0 - 17.8 \\ 
0.3 &  1.0 - 10.0 \\ 
0.2 &  1.0 - 5.6 \\ 
0.1 &  1.0 - 3.2 \\ 
0.0 & 1.0 - 3.2 \\ 

\tableline
\end{tabular}
\end{center}
\caption[]{\label{Table:collapCon} 
This table gives the non-rejectable contrast values contained in the 3-$\sigma$ contours of Figure~\ref{fig:contour}, arranged by faint slope, $\alpha_f$.
A contrast value of 1 corresponds to a knee and all other contrast values are divot $H$-distributions.
}
\end{table}

\begin{table}
\begin{center}
\begin{tabular}{   *{3}{c}  }
\hline
\hline

\multicolumn{3}{c}{\bf{Divots}} \\
\bf{$\alpha_f$} & \bf{contrast} &  \bf{ Pop. Est. } \\ 
\bf{} & \bf{} & \bf{\# [$H < 12$]} \\
\hline

0.4 & 1.8 & 6.4$\times10^{5}$\\
0.5 & 3.2 & 7.6$\times 10^{5}$\\
0.5 & 5.6 & 5.0$\times 10^{5}$\\
0.6 & 5.6 & 8.3$\times 10^{5}$\\
0.6 & 10 & 5.9$\times 10^{5}$\\

\hline
\multicolumn{3}{c}{ \bf{Knees}} \\
\bf{$\alpha_f$} & & \bf{ Pop. Est. } \\ 
\bf{} & & \bf{\# [$H < 12$]} \\
\hline

0.4 &  & 7.9$\times 10^{5}$ \\
0.3 &  & 3.6$\times 10^{5}$ \\
0.2 &  & 5.3$\times 10^{5}$ \\
0.1 &  & 2.4$\times 10^{5}$ \\

\tableline
\end{tabular}
\end{center}
\caption[]{\label{Table:popEst} 
Population estimates for divots along the $1-\sigma$ ridge in Figure~\ref{fig:contour}  and all knees that are not rejectable at the 1-$\sigma$ level. 
Population estimates are given to $H_r$ = 12, the limit of the observed sample.
Population estimates are determined by counting the number of required model object draws for a  simulated surveys to reproduce the observed number of detections, 22.}
\end{table}


\section{Discussion}
\label{discussion}

We have constrained the form of the scattering TNO $H$-distribution to a set of acceptable $\alpha_f$ and $c$ pairs. 
In this section we discuss the choice of model, including the orbital model and color distribution, and we argue in favor of a divot $H$-distribution for the scattering TNOs.

\subsection{Choice of Model}
\label{modelChoice}

\begin{figure}[h]
\includegraphics[width=1.2\textwidth]{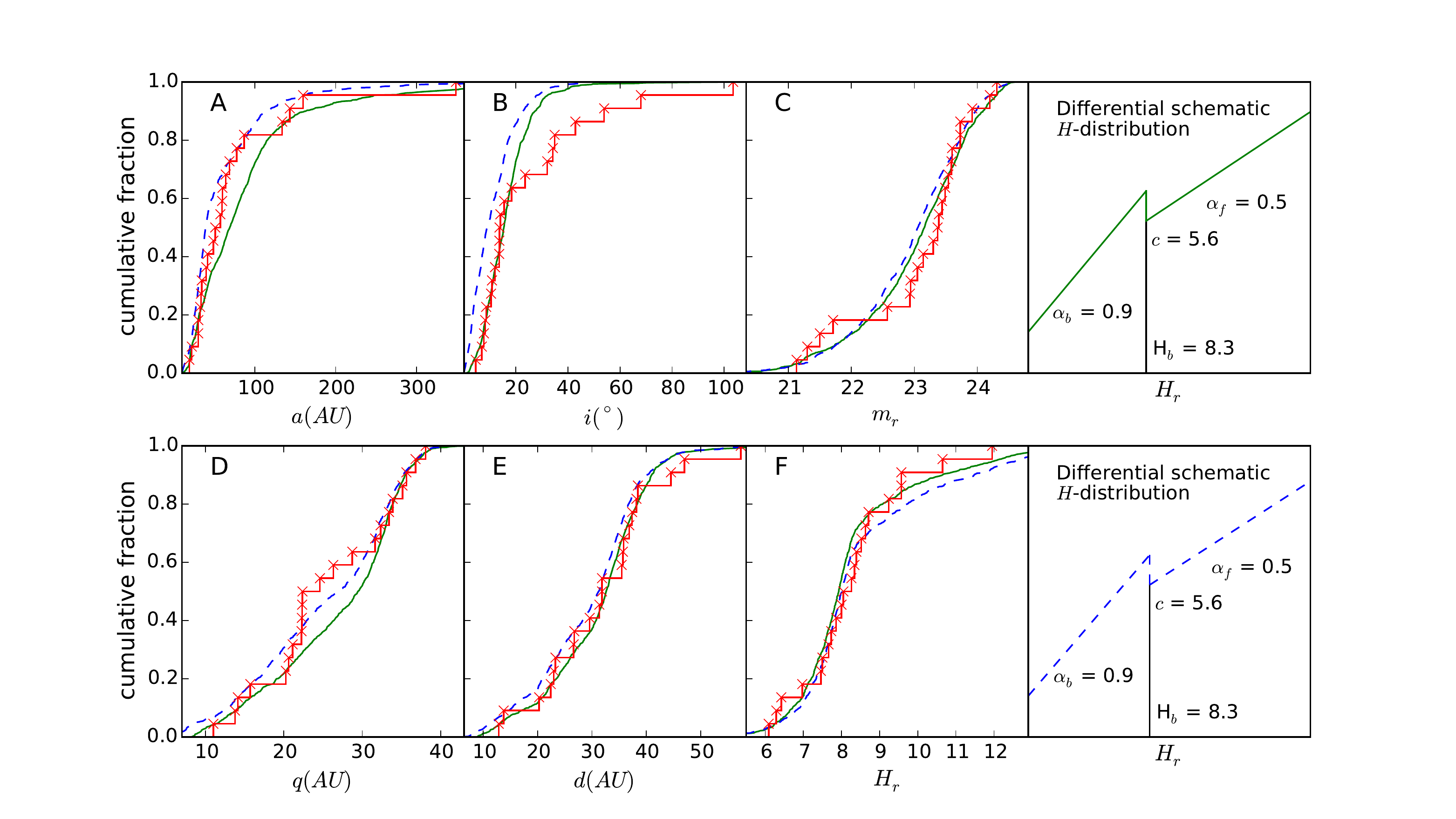}
\caption{ 
Cumulative distributions for the hotter inclination (solid green) and colder inclination (dashed blue) KRQ11 models using the same $H$-distribution for clear comparison. 
Colors and panels are as in Figure~\ref{fig:cumulativeDSS} with the exclusion of the intrinsic orbital model distributions (black dash-dot in Figure~\ref{fig:cumulativeDSS}).
In panel A and in panel B the two KRQ11 orbital model distributions (cold, hot) differ strongly from each other. 
Panel A shows that the $a$-distribution is strongly dependent on the initial $i$-distribution. 
In panels C-F the two KRQ11 orbital models (cold, hot) only show minor variations from each other; testing with either model on these distributions does not affect the conclusions of this work.
}
\label{fig:cumulativeHC}
\end{figure}

\begin{figure}[h]
\includegraphics[width=1\textwidth]{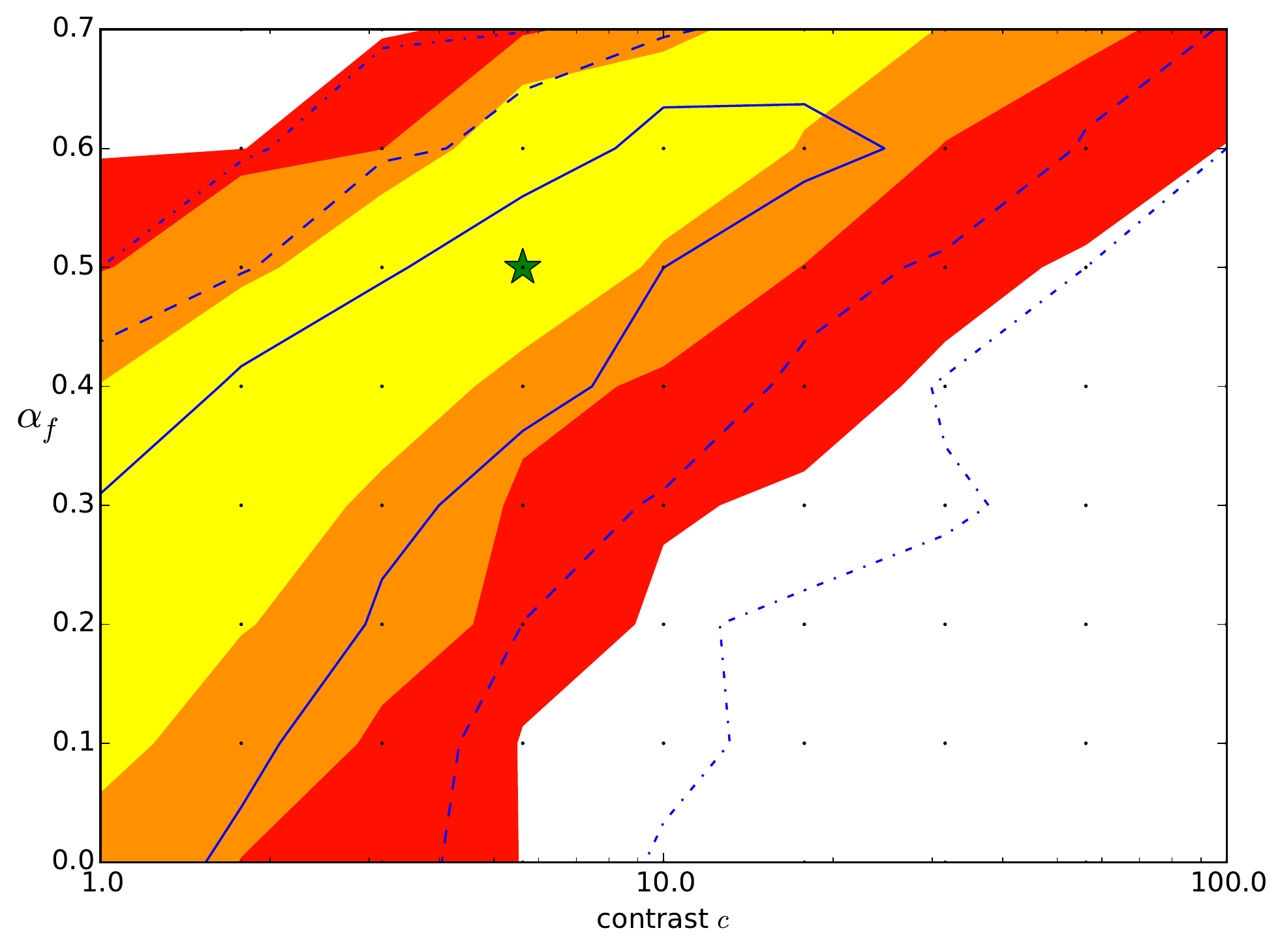}
\caption{ 
Comparison of the contours of rejectability for the hotter KRQ11 orbital model (colors as in Figure~\ref{fig:contour}) and the colder inclination KRQ11 orbital model (blue overlaid lines).
The blue lines represent the 1-$\sigma$ (solid), 2-$\sigma$ (dashed) and 3-$\sigma$ (dash-dot) contours for the colder KRQ11 orbital model. 
The green star indicates our preferred  ($\alpha_f$, $c$) parameters. 
There is only minor tension between these models.
  }
\label{fig:contourHC}
\end{figure}

As this work is model dependent, it is important to examine the effects of model choice.
The scattering TNOs are insensitive to the history of the exact number and configuration of planets in the outer Solar System \citep{shankman13} as long as the end-state is the current planet configuration; the scattering TNO interactions are so disruptive and chaotic that the exact history is ``erased".
We have shown above that the KRQ11 $i$-distribution is discrepant with our observations, and we thus test the effects of the $i$-distribution on this work.
We perform the same analysis as above on the original version of the KRQ11 orbital model, which has a colder initial $i$-distribution that is then evolved forward for the age of the Solar System. 
Figure~\ref{fig:cumulativeHC} shows the simulator-observed cumulative distributions for both the initially hot (green) and initially cold (blue) models with the same $H$ and color distributions. 
While the $i$ and $a$ distributions vary dramatically between the two orbital models, the $m$, $q$, $d$ and $H$ distributions only show small variations that do not statistically differentiate the two models. 
Figure~\ref{fig:contourHC} shows the confidence contours for the hot model (solid contour as in Figure~\ref{fig:contour}) and cold model (blue line overlays) when testing the AD statistic on the $m$, $q$, $d$ and $H$ distributions.
There is only minor tension between these two models which have good agreement at the 3-$\sigma$ level.
The observed objects Drac (peculiar $i$) and MS9 (peculiar $a$) have distinct orbital elements, but are classified as scattering TNOs by the \citet{gladman08} criterion. 
Given their unusual orbital elements, we consider that they may not be members of the scattering TNO population and we examine the effect of their inclusion in the analysis.
Drac and MS9 have no effect on the conclusions of the analysis for the $a$ and $i$ distributions.
Removing Drac and MS9 from the analysis also removes the tension between the hot and cold orbital models when testing on $m$, $q$, $d$ and $H$ and has no effect on the conclusions of this work regarding the $a$ and $i$ distributions.  
Our analysis is not strongly dependent on the $i$-distribution of the orbital model when testing on the $m$, $q$, $d$ and $H$ distributions.


\begin{figure}[h]
\includegraphics[width=1\textwidth]{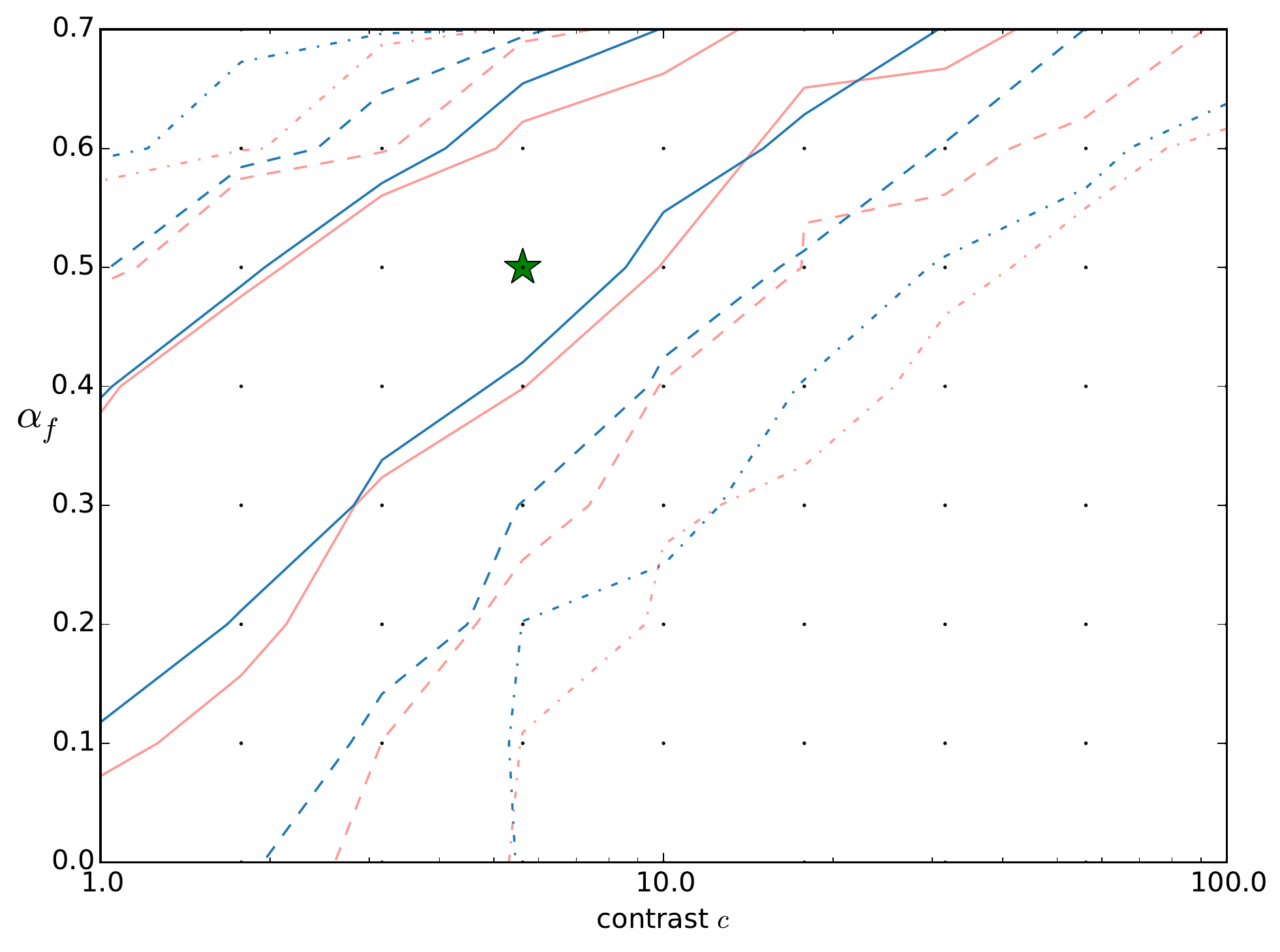}
\caption{1, 2, and 3-$\sigma$ contours (solid, dashed, and dash-dot lines respective) for two extreme cases of $g-r$ color choice. 
The blue lines denote $g-r = 0.5$ and the red lines denote $g-r$ = 0.9. 
The green star marks our preferred ($\alpha_f$, $c$) parameters.
There is no tension between these two color choices. 
}
\label{fig:contourCol}
\end{figure}

We also test the effects of the $g-r$ color distribution used.
In our analysis we draw model object $g-r$ colors from a gaussian centered at 0.7 with a standard deviation of 0.2, based on the measured CFEPS $g-r$ colors \citep{petit11}. 
We test the effects of our adopted $g-r$ distribution by performing the same analysis with fixed $g-r$ values at the 1-$\sigma$ extremes of our adopted distribution. 
Figure~\ref{fig:contourCol} shows the results of performing the same analysis with color conversions of $g-r$ = 0.5 and $g-r$ = 0.9.
There is no tension between rejectable ($\alpha_f$, $c$) parameters for these two very different  $g-r$ color values.
This analysis is not sensitive to exact knowledge of the intrinsic color of the TNOs in our sample, within reasonable bounds on those colors.

We test a combination of the orbital model, the color distribution and the $H$-distribution. 
We have shown that our analysis is not strongly dependent on the choice of color distribution nor on orbital model and thus when a candidate model is rejected, it is a rejection of the modeled $H$-distribution.


\subsection{Arguments for a divot}
\label{divotArguments}

Motivated by the measured JFC slopes (see \citet{solontoi12} Table 6), the Centaur slopes \citep{adams14}, and the theoretical slope of collisionally ground populations \citep{obrien05}, all of which suggest $\alpha_f \sim$ 0.5, we favor $\alpha_f = 0.5$. 
Our analysis has constrained the scattering TNO $H$-distribution to contrast values ranging from 1 (knees) to $\sim32$ (divots) for $\alpha_f = 0.5$. 
Here we consider the implications of a shared $H$-distribution among the hot populations of the outer Solar System.
We use the observed Neptune Trojan and Plutino samples and the JFC source problem to inform the scattering TNO $c$ value.  

Requiring the scattering TNOs to be the source of the JFCs sets strong lower limits on the size of the scattering TNO population \citep{volkmalhotra08}, which can be used to constrain the form of the $H$-distribution. 
\citet{volkmalhotra08} find $(0.8-1.7)\times10^{8}$ $D >$ 1 km scattering TNOs in the 30-50 AU heliocentric distance range are required in order for the scattering TNOs to be the JFC source region. 
To apply this constraint to our measured $H$-distribution values, we extend our knee and divot population estimates down to $D \sim \, 1$ km, using albedos of 5\% and 15\%, the range of measured albedos for the scattering TNOs and the Centaurs \citep{santossanz12, duffard14}.
$H$-distributions with $c \leq 10$ and $\alpha_f$ = 0.5 produce a scattering TNO population that is large enough to act as the source of the JFCs for both 5\% and 15\% albedos.
For $c=1$ (knees), our sample requires $\alpha_f \geq 0.4$ for 5\% albedo and $\alpha_f \geq 0.5$ for 15\% albedo.
If the $H$ distribution transitions to a shallow slope of $\alpha_f \sim 0.2$ as seen in \citet{fraser14} and not rejected by this analysis, it must then transition to a third, steeper, slope, $\alpha_{f2}$, at a smaller $H$ in order for the scattering TNOs to be numerous enough to act as the source population of the JFCs.
A transition to $\alpha_f \sim 0.2$ in the scattering TNO $H$-distribution requires two transitions, the faint one at a break that has yet to be observed, and three slopes in order to explain the source of the JFCs. 
The $H$-distribution of the scattering TNOs must transition to $\alpha_f \geq 0.4$ slope to provide a large enough scattering TNO population to act as the source for the JFCs.

Under the hypothesis that all of the hot TNO populations were implanted from the same source, a contrast value of $c > 1$ (a divot) is preferred as this would simultaneously explain the Neptune Trojan and Plutino $H$-distributions.
The Neptune Trojans must have been captured from the scattering TNOs and thus share the same $H$-distribution.
The observed lack of small-sized Neptune Trojans \citep{sheppardtrujillo10b, parker13} requires that there be a sharp decrease in the number of Neptune Trojans below a certain size.
Figure 3 of \citet{sheppardtrujillo10b} shows that this decrease must be in the form of a drop in the {\em differential} number distribution of Neptune Trojans in order to explain the lack of detected Neptune Trojans within their magnitude limit.
This drop can either be explained by a {\em differential} $H$-distribution of a transition to a steep negative slope, disfavoured by our sample of scattering TNOs, or an $H$-distribution with $c > 1$.
The preference of $c > 1$ from jointly considering the scattering TNOs and Neptune Trojans must also be consistent with the Plutino population if these three hot populations share a common source.
\citet{alexandersen15} find a decrease in observed Plutinos at $H$ magnitudes fainter (smaller) than our reported transitions in the scattering and Neptune Trojan $H$-distributions.
Exploring a variety of $H$-distribution parameters, \citet{alexandersen15} find that the divot parameters favored by \citet{shankman13}, with no tuning, provide a good representation of the observed $H$-distribution for $5 < H < 11$ Plutinos.
 A divot ($ c > 1 $) solution simultaneously matches the observed scattering TNO, Neptune Trojan, Centaur, and Plutino $H$-distributions.

\begin{figure}[h!]
\includegraphics[width=0.75\textwidth]{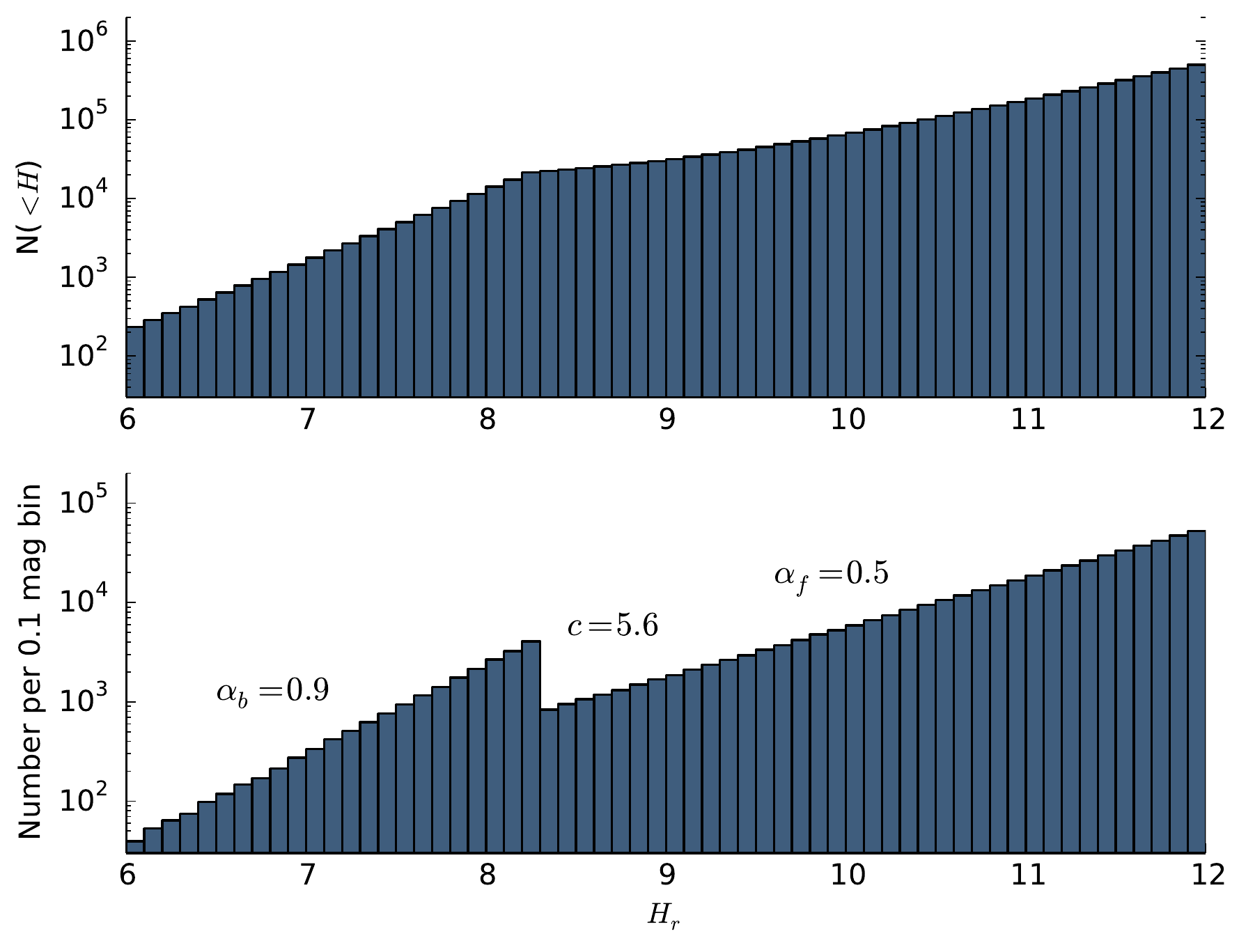}
\caption{
Histograms of the population estimate for our preferred ($\alpha_f$, $c$) $H$-distribution. The vertical axes show the estimated total scattering TNO population numbers for this $H$-distribution.
\textbf{Top:} The cumulative distribution. 
The divot with a contrast of $\sim 5.6$ is a striking feature when viewed differentially, but when viewed cumulatively only results in a small flattened section before $\alpha_f = 0.5$ dominates the distribution and the signature of a divot is erased.
\textbf{Bottom:} The differential distribution with 0.1 magnitude bins.
}
\label{fig:popEst}
\end{figure}

The \citet{shankman13} preferred $H$-distribution, with parameters of $\alpha_f = 0.5$ and $c \sim 5.6$, continues to provide an excellent match to our observations, even with the inclusion of two new surveys that add survey depth and double the observed sample. 
This $H$-distribution remains our preferred choice for its ability to match our observed Scattering TNO sample.
Figure~\ref{fig:popEst} shows the population estimates for our preferred divot in both cumulative and differential representations.
We provide the cumulative representation to highlight how such a visually obvious feature is completely suppressed when viewed cumulatively.
The flattening of the cumulative distribution seen in Figure 2 of \citet{sheppardtrujillo10b} that then gradually transitions to a positive slope is characteristic of a divot. Figure~\ref{fig:popEst} demonstrates that when you view a divot cumulatively, it produces exactly this signature.
If there is a divot in the $H$-distribution of the TNO populations, prior studies of the luminosity function, or even $H$, could be suppressing the signature of this feature if they combine surveys which were not characterized in the same way

A divot can also explain the rollover to negative slopes that some groups have measured (\citealt{bernstein04}; \citealt{fraserkavelaars09}). 
The transition in the size-frequency distribution for a divot would likely take the form of a steep negative slope over a short size range, rather than a discontinuous drop. 
If a survey terminates just past the break point, one could measure a signature of this transition as a negative slope without measuring the recovery.

Two different formation mechanisms have been shown to produce divot size-frequency distributions. 
\citet{fraser09} shows that a kink in the size-frequency distribution collisionally propagates to larger sizes and produces a divot. 
\citet{campobagatin12} explore the ``born big" scenario \citep{morbidelli09} in the Kuiper Belt, collisionally evolving a population with no initial objects smaller than 100 km. 
As the population evolves under collisions, a divot is produced and the small-sized objects have a collisional equilibrium slope and a contrast of the same scale ($\sim5$; see Fig. 6 of \citealt{campobagatin12}) as our preferred model parameters ($\alpha_f = 0.5$ and $c \sim 5.6$).
Divot size-frequency distributions are consistent with our understanding of small-body formation physics, and have been shown to arise from two different formation mechanisms.

Divot $H$-distributions are plausible, explain our observations and when generalized to other hot TNO populations can explain several outstanding questions.
Our preferred $H$-distribution parameter set is one of the least rejectable $H$-distributions we tested, provides a good match with the observed $H$-distribution (see Figure~\ref{fig:cumulativeSPL05}), is consistent with the observed JFC slopes, can solve the problem of JFC supply, is consistent with the observed Centaur $H$-distribution, is consistent with the observed Neptune Trojan and Plutino $H$-distributions and is consistent with divots produced in collisional models. 
We find a divot of contrast c $\sim 5.6$ and faint slope $\alpha_f$ to be compelling and argue that it provides a better, versus knees, solution to the observed lack of small-sized hot TNOs while simultaneously explaining the supply of JFCs.

\section{Conclusion}
\label{conclusion}

Using our 22 object scattering TNO sample, we have constrained the form of the scattering TNO $H$-distribution as a proxy for its size-frequency distribution. 
Our sample rules out all single-slope distributions and constrains the form to divots or shallow ($\alpha_f < 0.6$) knees.
We find that a particular scattering orbital model for the formation of the scattering TNOs is not consistent with the observed $i$ distribution, producing too few high-$i$ scatterers, possibly indicating an additional inclination component in the scattering TNOs.
We argue in favor of a divot $H$-distribution and find a population estimate of (2.4-8.3)$\times 10^5$ down to $H_r = 12$ is required to match our observed sample.
Extrapolating our preferred divot $H$-distribution down to the km scale predicts a scattering TNO population large enough to act as the sole supply of the JFCs, allowing a divot to explain JFC supply, while simultaneously explaining the observed Neptune Trojan, Plutino and scattering TNO size or $H$ distributions.

\acknowledgments 
This project was funded by the National Science and Engineering Research Council and the National Research Council of Canada. This research used the facilities of the Canadian Astronomy Data Centre operated by the National Research Council of Canada with the support of the Canadian Space Agency.


{\it Facilities:} \facility{CFHT}, \facility{NRC}, \facility{CANFAR}.



\appendix






\bibliographystyle{apj}
\bibliography{bibthesis}

\begin{thebibliography}{}
\expandafter\ifx\csname natexlab\endcsname\relax\def\natexlab#1{#1}\fi

\bibitem[{{Adams} {et~al.}(2014){Adams}, {Gulbis}, {Elliot}, {Benecchi},
  {Buie}, {Trilling}, \& {Wasserman}}]{adams14}
{Adams}, E.~R., {Gulbis}, A.~A.~S., {Elliot}, J.~L., {et~al.} 2014, \aj, 148,
  55

\bibitem[{{Alexandersen} {et~al.}(2013){Alexandersen}, {Gladman},
  {Greenstreet}, {Kavelaars}, {Petit}, \& {Gwyn}}]{alexandersen13}
{Alexandersen}, M., {Gladman}, B., {Greenstreet}, S., {et~al.} 2013, Science,
  341, 994

\bibitem[{{Alexandersen} {et~al.}(2014){Alexandersen}, {Gladman}, {Kavelaars},
  {Petit}, {Gwyn}, \& {Shankman}}]{alexandersen15}
{Alexandersen}, M., {Gladman}, B., {Kavelaars}, J.~J., {et~al.} 2014, ArXiv
  e-prints, arXiv:1411.7953

\bibitem[{Anderson \& Darling(1954)}]{andersondarling54}
Anderson, T.~W., \& Darling, D.~A. 1954, Journal of the American Statistical
  Association, 49, 765

\bibitem[{{Bannister} {et~al.}(2016){Bannister}, {Kavelaars}, {Gladman}, \&
  {Petit}}]{bannister15}
{Bannister}, M.~T., {Kavelaars}, J.~J., {Gladman}, B.~J., \& {Petit}, J.-M.
  2016, AJ, submitted

\bibitem[{{Batygin} \& {Brown}(2010)}]{batyginbrown10}
{Batygin}, K., \& {Brown}, M.~E. 2010, \apj, 716, 1323

\bibitem[{{Benecchi} \& {Sheppard}(2013)}]{benecchisheppard13}
{Benecchi}, S.~D., \& {Sheppard}, S.~S. 2013, \aj, 145, 124

\bibitem[{{Bernstein} {et~al.}(2004){Bernstein}, {Trilling}, {Allen}, {Brown},
  {Holman}, \& {Malhotra}}]{bernstein04}
{Bernstein}, G.~M., {Trilling}, D.~E., {Allen}, R.~L., {et~al.} 2004, \aj, 128,
  1364

\bibitem[{{Brown}(2001)}]{brown01}
{Brown}, M.~E. 2001, \aj, 121, 2804

\bibitem[{{Campo Bagatin} \& {Benavidez}(2012)}]{campobagatin12}
{Campo Bagatin}, A., \& {Benavidez}, P.~G. 2012, \mnras, 423, 1254

\bibitem[{{Duffard} {et~al.}(2014){Duffard}, {Pinilla-Alonso}, {Santos-Sanz},
  {Vilenius}, {Ortiz}, {Mueller}, {Fornasier}, {Lellouch}, {Mommert}, {Pal},
  {Kiss}, {Mueller}, {Stansberry}, {Delsanti}, {Peixinho}, \&
  {Trilling}}]{duffard14}
{Duffard}, R., {Pinilla-Alonso}, N., {Santos-Sanz}, P., {et~al.} 2014, \aap,
  564, A92

\bibitem[{{Fraser}(2009)}]{fraser09}
{Fraser}, W.~C. 2009, \apj, 706, 119

\bibitem[{{Fraser} {et~al.}(2014){Fraser}, {Brown}, {Morbidelli}, {Parker}, \&
  {Batygin}}]{fraser14}
{Fraser}, W.~C., {Brown}, M.~E., {Morbidelli}, A., {Parker}, A., \& {Batygin},
  K. 2014, \apj, 782, 100

\bibitem[{{Fraser} {et~al.}(2010){Fraser}, {Brown}, \& {Schwamb}}]{fraser10}
{Fraser}, W.~C., {Brown}, M.~E., \& {Schwamb}, M.~E. 2010, \icarus, 210, 944

\bibitem[{{Fraser} \& {Kavelaars}(2008)}]{fraserkavelaars08}
{Fraser}, W.~C., \& {Kavelaars}, J.~J. 2008, \icarus, 198, 452

\bibitem[{{Fraser} \& {Kavelaars}(2009)}]{fraserkavelaars09}
---. 2009, \aj, 137, 72

\bibitem[{{Fuentes} \& {Holman}(2008)}]{fuentesholman08}
{Fuentes}, C.~I., \& {Holman}, M.~J. 2008, \aj, 136, 83

\bibitem[{{Gladman} \& {Kavelaars}(1997)}]{gladmankavelaars97}
{Gladman}, B., \& {Kavelaars}, J.~J. 1997, \aap, 317, L35

\bibitem[{{Gladman} {et~al.}(2001){Gladman}, {Kavelaars}, {Petit},
  {Morbidelli}, {Holman}, \& {Loredo}}]{gladman01}
{Gladman}, B., {Kavelaars}, J.~J., {Petit}, J.-M., {et~al.} 2001, \aj, 122,
  1051

\bibitem[{{Gladman} {et~al.}(2008){Gladman}, {Marsden}, \&
  {Vanlaerhoven}}]{gladman08}
{Gladman}, B., {Marsden}, B.~G., \& {Vanlaerhoven}, C. 2008, {Nomenclature in
  the Outer Solar System} (The University of Arizona Press), 43--57

\bibitem[{{Gladman} {et~al.}(2012){Gladman}, {Lawler}, {Petit}, {Kavelaars},
  {Jones}, {Parker}, {Van Laerhoven}, {Nicholson}, {Rousselot}, {Bieryla}, \&
  {Ashby}}]{gladman12}
{Gladman}, B., {Lawler}, S.~M., {Petit}, J.-M., {et~al.} 2012, Astronomical
  Journal, 144, 23

\bibitem[{{Gomes}(2003)}]{gomes03}
{Gomes}, R.~S. 2003, \icarus, 161, 404

\bibitem[{{Jewitt} {et~al.}(1998){Jewitt}, {Luu}, \& {Trujillo}}]{jewitt98}
{Jewitt}, D., {Luu}, J., \& {Trujillo}, C. 1998, \aj, 115, 2125

\bibitem[{{Jones} {et~al.}(2006){Jones}, {Gladman}, {Petit}, {Rousselot},
  {Mousis}, {Kavelaars}, {Campo Bagatin}, {Bernabeu}, {Benavidez}, {Parker},
  {Nicholson}, {Holman}, {Grav}, {Doressoundiram}, {Veillet}, {Scholl}, \&
  {Mars}}]{jones06}
{Jones}, R.~L., {Gladman}, B., {Petit}, J.-M., {et~al.} 2006, \icarus, 185, 508

\bibitem[{{Kaib} {et~al.}(2011){Kaib}, {Ro{\v s}kar}, \& {Quinn}}]{kaib11b}
{Kaib}, N.~A., {Ro{\v s}kar}, R., \& {Quinn}, T. 2011, Icarus, 215, 491

\bibitem[{{Kavelaars} {et~al.}(2008){Kavelaars}, {Jones}, {Gladman}, {Parker},
  \& {Petit}}]{kavelaars08}
{Kavelaars}, J., {Jones}, L., {Gladman}, B., {Parker}, J.~W., \& {Petit}, J.-M.
  2008, {The Orbital and Spatial Distribution of the Kuiper Belt} (The
  University of Arizona Press), 59--69

\bibitem[{{Kavelaars} {et~al.}(2011){Kavelaars}, {Petit}, {Gladman}, {Jone},
  {Parker}, \& {Taylor}}]{kavelaarsepsc11}
{Kavelaars}, J.~J., {Petit}, J.-M., {Gladman}, B., {et~al.} 2011, in EPSC-DPS
  Joint Meeting 2011, 1318

\bibitem[{{Leinhardt} {et~al.}(2008){Leinhardt}, {Stewart}, \&
  {Schultz}}]{Leinhardt2007}
{Leinhardt}, Z.~M., {Stewart}, S.~T., \& {Schultz}, P.~H. 2008, {Physical
  Effects of Collisions in the Kuiper Belt} (The University of Arizona Press),
  195--211

\bibitem[{{Levison} \& {Duncan}(1994)}]{levisonduncan94}
{Levison}, H.~F., \& {Duncan}, M.~J. 1994, Icarus, 108, 18

\bibitem[{{Levison} {et~al.}(2008){Levison}, {Morbidelli}, {Van Laerhoven},
  {Gomes}, \& {Tsiganis}}]{levison08}
{Levison}, H.~F., {Morbidelli}, A., {Van Laerhoven}, C., {Gomes}, R., \&
  {Tsiganis}, K. 2008, \icarus, 196, 258

\bibitem[{{Morbidelli} {et~al.}(2009){Morbidelli}, {Bottke}, {Nesvorn{\'y}}, \&
  {Levison}}]{morbidelli09}
{Morbidelli}, A., {Bottke}, W.~F., {Nesvorn{\'y}}, D., \& {Levison}, H.~F.
  2009, \icarus, 204, 558

\bibitem[{{Nesvorny}(2015)}]{nesvorny15}
{Nesvorny}, D. 2015, ArXiv e-prints, arXiv:1504.06021

\bibitem[{{O'Brien} \& {Greenberg}(2005)}]{obrien05}
{O'Brien}, D.~P., \& {Greenberg}, R. 2005, \icarus, 178, 179

\bibitem[{{Parker}(2015)}]{parker15}
{Parker}, A.~H. 2015, \icarus, 247, 112

\bibitem[{{Parker} {et~al.}(2013){Parker}, {Buie}, {Osip}, {Gwyn}, {Holman},
  {Borncamp}, {Spencer}, {Benecchi}, {Binzel}, {DeMeo}, {Fabbro}, {Fuentes},
  {Gay}, {Kavelaars}, {Stern}, {Tholen}, {Trilling}, {Ragozzine}, {Wasserman},
  \& {Ice Hunters}}]{parker13}
{Parker}, A.~H., {Buie}, M.~W., {Osip}, D.~J., {et~al.} 2013, Astronomical
  Journal, 145, 96

\bibitem[{{Petit} {et~al.}(2008){Petit}, {Kavelaars}, {Gladman}, \&
  {Loredo}}]{petit08}
{Petit}, J.-M., {Kavelaars}, J.~J., {Gladman}, B., \& {Loredo}, T. 2008, {Size
  Distribution of Multikilometer Transneptunian Objects} (The University of
  Arizona Press), 71--87

\bibitem[{{Petit} {et~al.}(2015){Petit}, {Kavelaars}, {Gladman}, {Jones}, \&
  {Parker}}]{petit15}
{Petit}, J.-M., {Kavelaars}, J.~J., {Gladman}, B.~J., {Jones}, R.~L., \&
  {Parker}, J.~W. 2015, AJ, inprep

\bibitem[{{Petit} {et~al.}(2011){Petit}, {Kavelaars}, {Gladman}, {Jones},
  {Parker}, {Van Laerhoven}, {Nicholson}, {Mars}, {Rousselot}, {Mousis},
  {Marsden}, {Bieryla}, {Taylor}, {Ashby}, {Benavidez}, {Campo Bagatin}, \&
  {Bernabeu}}]{petit11}
{Petit}, J.-M., {Kavelaars}, J.~J., {Gladman}, B.~J., {et~al.} 2011,
  Astronomical Journal, 142, 131

\bibitem[{{Santos-Sanz} {et~al.}(2012){Santos-Sanz}, {Lellouch}, {Fornasier},
  {Kiss}, {Pal}, {M{\"u}ller}, {Vilenius}, {Stansberry}, {Mommert}, {Delsanti},
  {Mueller}, {Peixinho}, {Henry}, {Ortiz}, {Thirouin}, {Protopapa}, {Duffard},
  {Szalai}, {Lim}, {Ejeta}, {Hartogh}, {Harris}, \& {Rengel}}]{santossanz12}
{Santos-Sanz}, P., {Lellouch}, E., {Fornasier}, S., {et~al.} 2012, \aap, 541,
  A92

\bibitem[{{Shankman}(2012)}]{shankman12}
{Shankman}, C. 2012, Master's thesis, University of British Columbia

\bibitem[{{Shankman} {et~al.}(2013){Shankman}, {Gladman}, {Kaib}, {Kavelaars},
  \& {Petit}}]{shankman13}
{Shankman}, C., {Gladman}, B.~J., {Kaib}, N., {Kavelaars}, J.~J., \& {Petit},
  J.~M. 2013, Astrophysical Journal Letters, 764, L2

\bibitem[{{Sheppard} \& {Trujillo}(2010)}]{sheppardtrujillo10b}
{Sheppard}, S.~S., \& {Trujillo}, C.~A. 2010, Astrophysical Journal Letters,
  723, L233

\bibitem[{{Solontoi} {et~al.}(2012){Solontoi}, {Ivezi{\'c}}, {Juri{\'c}},
  {Becker}, {Jones}, {West}, {Kent}, {Lupton}, {Claire}, {Knapp}, {Quinn},
  {Gunn}, \& {Schneider}}]{solontoi12}
{Solontoi}, M., {Ivezi{\'c}}, {\v Z}., {Juri{\'c}}, M., {et~al.} 2012, \icarus,
  218, 571

\bibitem[{{Stansberry} {et~al.}(2008){Stansberry}, {Grundy}, {Brown},
  {Cruikshank}, {Spencer}, {Trilling}, \& {Margot}}]{stansberry08}
{Stansberry}, J., {Grundy}, W., {Brown}, M., {et~al.} 2008, {Physical
  Properties of Kuiper Belt and Centaur Objects: Constraints from the Spitzer
  Space Telescope} (The University of Arizona Press), 161--179

\bibitem[{{Trujillo} {et~al.}(2000){Trujillo}, {Jewitt}, \& {Luu}}]{trujillo00}
{Trujillo}, C.~A., {Jewitt}, D.~C., \& {Luu}, J.~X. 2000, \apjl, 529, L103

\bibitem[{{Volk} \& {Malhotra}(2008)}]{volkmalhotra08}
{Volk}, K., \& {Malhotra}, R. 2008, Astrophysical Journal, 687, 714

\end{thebibliography}



\clearpage


\end{document}